\documentclass[aps,prl,showpacs,twocolumn,amsmath,amssymb,superscriptaddress]{revtex4}

\usepackage[colorlinks,citecolor=blue,linkcolor=cyan]{hyperref}
\usepackage[usenames,dvipsnames,svgnames,table]{xcolor}

\usepackage{graphicx}
\usepackage{Jonasmacros}
\usepackage{bm}
\usepackage{mathptmx}
\usepackage{txfonts}
\usepackage{times}

\renewcommand{\section}[1]{{\par\it #1.---}\ignorespaces}

\begin{document}
\date{\today}
\title{Filling of magnetic-impurity-induced gap in topological insulators by potential scattering}

\author{A. M. Black-Schaffer}
\email{annica.black-schaffer@physics.uu.se}
\affiliation{Department of Physics and Astronomy, Uppsala University, Box 530, SE-751 21\ Uppsala, Sweden}

\author{A. V. Balatsky}
\affiliation{NORDITA, Roslagstullsbacken 23, SE-106 91\ \ Stockholm, Sweden}
\affiliation{Institute for Materials Science, Los Alamos, New Mexico 87545, USA}

\author{J. Fransson}
\email{jonas.fransson@physics.uu.se}
\affiliation{Department of Physics and Astronomy, Uppsala University, Box 530, SE-751 21\  Uppsala, Sweden}

\begin{abstract}
We show that the energy gap induced by ferromagnetically aligned magnetic impurities on the surface of a topological insulator can be filled, due to scattering off the non-magnetic potential of the impurities. In both a continuum surface model and a three-dimensional tight-binding lattice model, we find that the energy gap disappears already at weak potential scattering as impurity resonances add spectral weight at the Dirac point. This can help explain seemingly contradictory experimental results as to the existence of a gap.   
\end{abstract}
\maketitle

The recent discovery of topological insulators (TIs) \cite{Kane_PRL05,Bernevig_2006, Koenig_Science07, Hsieh08} has led to exciting possibilities for producing electronically engineered states. TIs are bulk insulators but host metallic surface states with a Dirac-like dispersion \cite{Kane_RMP2010, Qi11RMP}. The momentum-spin locking in the surface states \cite{Kane_PRL07, Moore_PRB07, Qi08, Hsieh09spin} offers unique technological capabilities, especially if an energy gap can be created. 

A gap in the TI surface states can be generated by time-reversal breaking perturbations and one of the currently most discussed approaches for engineering a gap is magnetic doping \cite{Chen10, Wray11, Xu12,Lee03022015}. The idea is that ferromagnetically ordered impurities will produce a net magnetic field, which then gaps the TI surface states \cite{SCZhangimp, Abanin11}. Beyond an energy gap, this would also give rise to multiple other exotic phenomena, such as the anomalous Hall effect recently observed \cite{Qu13082010}.
However, despite a multitude of recent experimental studies, evidence for a gap in the TI surface states from magnetic doping remains controversial. Both angle-resolved photoemission spectroscopy (ARPES) and scanning tunneling spectroscopy (STS) measurements have reported the presence of an energy gap \cite{Chen10, Wray11, Xu12, Lee03022015}, while other similar studies have found no gap \cite{Scholz12, Valla12, Honolka12PRL, Schlenk13PRL, Sessi14}. Interestingly, several studies have even reported no significant difference between magnetic and non-magnetic surface impurities \cite{Bianchi11, Valla12}.

The key assumption, when expecting an energy gap from magnetic impurities, is that the TI surface electrons only see an average Zeeman magnetic field. Here we point out that magnetic impurities, such as Fe or Cr, also strongly scatter electrons. Thus, in addition to an effective magnetic field, the presence of magnetic impurities also adds non-magnetic, or potential, scattering. 
Potential scattering is known to induce low-energy impurity resonances in Dirac materials \cite{DMreview}, ranging from graphene \cite{Gomez-Rodriguez_PRL10} and $d$-wave superconductors \cite{Balatsky_RMP}, to TIs \cite{Black-Schaffer12imp, Teague12, Alpichshev_PRL12, JFransson14}.
 
In this work, we investigate the effect on the TI surface states of both the magnetic and potential scattering by magnetic impurities. 
We find that there are two effects that simultaneously modify the Dirac spectrum of the TI surface states. 
First, the presence of magnetic scattering opens a gap in the spectrum. The gap is generated due to the magnetic scattering $(M = JS_z)$ modifying the energy dispersion relation $E = \hbar v_Fk \rightarrow \sqrt{(\hbar v_F)^2 k^2 + M^2}$.
Secondly, the potential scattering ($U$) induces impurity resonance states, which add low-energy states to the spectrum. 
The relative strength of the potential and magnetic scattering determines the net density of states (DOS). In the case of weak potential scattering ($U \lesssim M$), we find a well-defined gap. On the other hand, for stronger potential scattering ($U\gtrsim M$), the tail of the impurity resonances dominates the low-energy spectrum around the Dirac point and the gap is filled. This provides a unifying framework that allows us to reconcile the conflicting claims about the presence/absence of a gap at the Dirac point in magnetically doped TIs.
More specifically, we use both a continuum surface model and a tight-binding three-dimensional (3D) lattice model of a TI, where we introduce dilute concentrations of magnetic impurities. We find that these two distinct models generate remarkably similar results and thus produce a convincing picture regarding the role of magnetic and potential contributions in magnetic impurity scattering.

\section{Continuum surface model}
We first consider an effective continuum model of the surface states of a TI and their coupling to local magnetic impurities using the Hamiltonian
\begin{align}
\Hamil_{\rm surf}=&
	\sum_\bfk \hbar v_F \psi^\dagger_\bfk (\bfk\times\hat{\bf z})\cdot\bfsigma\psi_\bfk
	+\int\psi^\dagger(\bfr)\bfV(\bfr)\psi(\bfr)d\bfr
	.
\label{eq-model}
\end{align}
Here $\hbar v_F$ is the Fermi velocity of the surface states, $\bfsigma$ denotes the vector of Pauli matrices, and $\psi_\bfk=(\psi_{\bfk\up} ,\ \psi_{\bfk\down})^t=\int\psi(\bfr)e^{-i\bfk\cdot\bfr}d\bfr$ is the electron annihilation spinor at momentum $\bfk$. The magnetic impurities are modeled by the total scattering potential $\bfV(\bfr)=\sum_m(U\sigma_0-J\bfS\cdot\bfsigma)\delta(\bfr-\bfr_m)$, which includes both scattering off a potential $U$, with the identity matrix $\sigma_0$, and magnetic moment $\bfS$, both acting as point defects at positions $\bfr_m$. For simplicity, since the quantum nature of the spins is not crucial, we use large spin moments $|\bfS|\rightarrow\infty$ and weak couplings $J\rightarrow0$, requiring $J|\bfS|=$ constant, such that the impurity spins can be treated as classical.

With the spin of the surface states oriented within the $xy$-plane in $\Hamil_{\rm surf}$, a magnetic field along the $\hat{\bf z}$-direction gaps the surface spectrum at the Dirac point, as it adds a term proportional to $\sigma_z$ \cite{Kane_RMP2010, Qi11RMP, DMreview}. 
Single magnetic impurities with moment along $\hat{\bf z}$ (with potential scattering ignored) and with a finite spatial extent have also been shown to give an effective local gap \cite{SCZhangimp}, while a single point-like magnetic impurity results in no incipient gap \cite{Biswas10}. For finite concentration of point-like magnetic impurities, their spins has been shown to align and thus produce an effective magnetic field \cite{Abanin11}, and we are primarily interested in this latter system.

To proceed, we study the scattering off a finite concentration of impurities in Eq.~\eqref{eq-model} using the $T$-matrix approach, see e.g.~Refs.~\cite{Peres08, JFransson14}.  For a low density of localized magnetic impurities, all with the scattering potential $\bfV(\bfr)$, the impurity averaged Fermion Green function (GF) reads $\bfG(\bfk,z)=\bfG_0(\bfk,z)[\sigma_0-\bfSigma(\bfk,z)\bfG_0(\bfk,z)]^{-1}$ to first order in the density of impurities $\rho$.  Here, the self-energy in the Born approximation is given by $\bfSigma(\bfk,z)=\rho[\sigma_0-\bfV_0g_0(z)]^{-1}\bfV_0$, with $\bfV_0=(U\sigma_0-J\bfS\cdot\bfsigma)$. Resolving the algebra, we obtain
\begin{subequations}
\label{eq-GBA}
\begin{align}
\bfG(\bfk,z)=&
	\frac{
		[z-\widetilde{U}(z)]\sigma_0+[\hbar v_F \bfk\times\hat{\bf z}+\widetilde\bfM(z)]\cdot\bfsigma
	}{
		[z-\widetilde{U}(z)]^2-\widetilde M_z(z)-|\hbar v_Fke^{i\varphi}+\widetilde M_+(z)|^2
	}
	,
\label{eq-Gk}
\\
\widetilde{U}(z)=&
	\rho\frac{U-g_0(z)[U^2-M^2]}{[1-g_0(z)U]^2-g_0^2(z)M^2}
,
\label{eq-U}
\\
\widetilde{\bfM}(z)=&
	\frac{\rho}{[1-g_0(z)U]^2-g_0^2(z)M^2}
	\bfM.
\label{eq-Delta}
\end{align}
\end{subequations}
Here $g_0(z)\sigma_0=\sum_\bfk\bfg(\bfk,z)=-z\log[D_c/(-z)]\sigma_0/4\pi(\hbar v_F)^2$, where we have introduced a finite cut-off energy $D_c$ for the band width of the surface states. In addition, we have introduced $\bfM=-J\bfS$, $M=|\bfM|$, $M_+=M_x+iM_y$, and $\tan\varphi=k_x/k_y$.

\begin{figure}[t]
\begin{center}
\includegraphics[width=0.99\columnwidth]{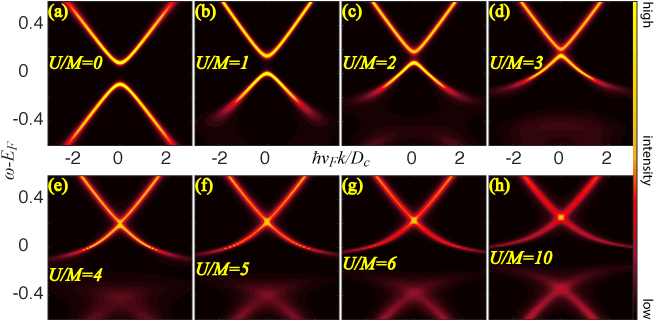}
\end{center}
\caption{(Color online) Evolution of the band structure as a function of the impurity potential $U/M = 0,\ 1,\ 2,\ 3,\ 4,\ 5,\ 6,\ 10$. Here, $\rho=0.1$, $\bfM=M\hat{\bf z}$, $D_c/\hbar v_F=300$, and $\hbar v_F=1$.
}
\label{fig-Nk}
\end{figure}
Using $-\im\bfG^r(\bfk,\omega)/\pi$ to calculate the low-energy band structure, the formulation provided in Eqs.~(\ref{eq-GBA}) enables a continuous variation of the potential scattering $U$. In Fig.~\ref{fig-Nk} we plot the evolution of the calculated band structure for increasing potential scattering $U$, but fixed strength of the local magnetic impurity moment $\bfM = M\hat{\bf z}$. We here explicitly choose a magnetic moment such that there is a clear gap at the Dirac point for zero potential scattering. The band structure shows an essentially unperturbed and gapped band structure for potentials $U/M\lesssim1$, see Figs.~\ref{fig-Nk}(a) and (b). However, the gap vanishes rapidly for increasing $U/M\gtrsim1$, as seen in Figs.~\ref{fig-Nk}(b) -- (h). 
The flattening of the lower part of the Dirac cone for $U/M\gtrsim1$ indicates the presence of an impurity resonance. Even though the impurity resonance is located well below the Dirac point, there is enough spectral weight available from its tail to significantly modify the gap and for $U/M\geq 4$ the gap is completely filled. We thus conclude that even weak potential scattering significantly alters the low-energy spectrum by filling the magnetically induced gap at the Dirac point. 
There is also an overall upward energy shift of the band structure with increasing $U$, indicated by the shift $z\rightarrow z-\widetilde{U}(z)$ in the denominator of $\bfG(\bfk,z)$ in Eqs.~\eqref{eq-GBA}. This is expected since potential scattering contributes as a local positive (hole) doping. This effect is advantageous since it both breaks particle-hole symmetry, not usually present in TIs \cite{Black-Schaffer12imp}, and it can also be seen to mimic additional carrier doping of the material. 
Both the gap turning into a region of suppressed, but finite, intensity and the energy shift in Fig.~\ref{fig-Nk} agree very well with the experimental results reported in Ref.~\cite{Chen10}.
The results in Fig.~\ref{fig-Nk} are for magnetic moments along $\sigma_z$. Adding a small moment also along $\sigma_{x,y}$ does not change the results. This is clear since the shift $\hbar v_F ke^{i\varphi} \rightarrow \hbar v_Fke^{i\varphi}+\widetilde{M}_+(z)$ in the dispersion relation in Eqs.~\ref{eq-GBA} does not act as a mass term in the Hamiltonian, but merely  renormalizes the angular dependence of the energy.

\begin{figure}[t]
\begin{center}
\includegraphics[width=0.99\columnwidth]{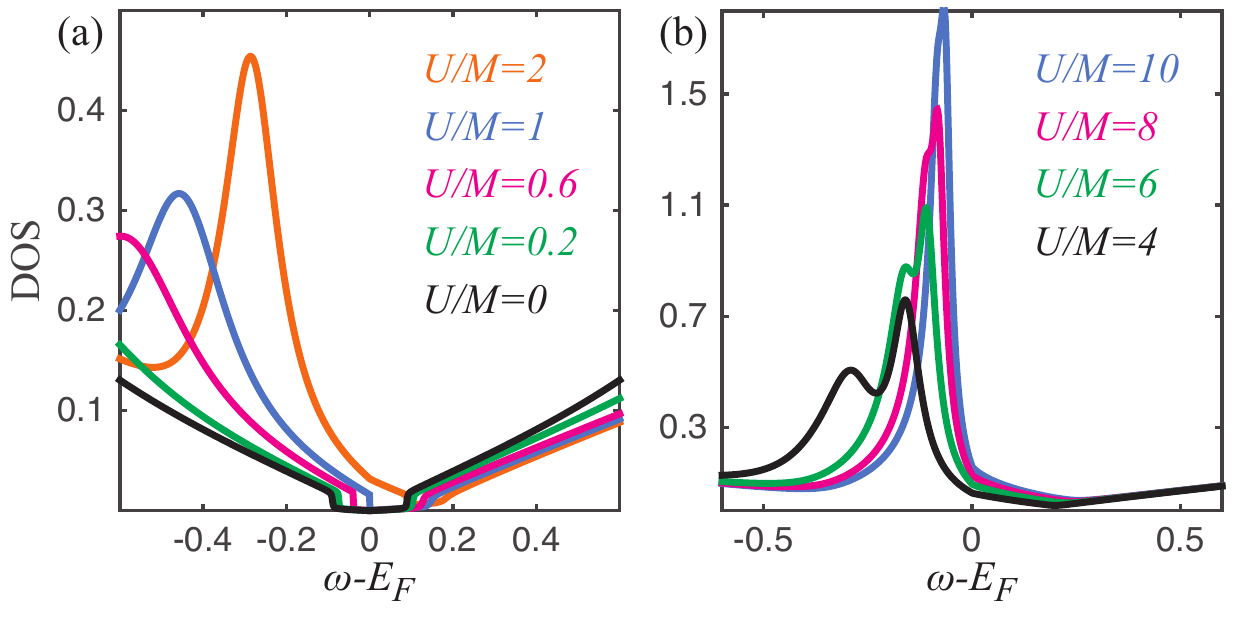}
\end{center}
\caption{(Color online) Evolution of the DOS as a function of the impurity potential $U/M= 0,\ 0.2,\ 0.6,\ 1,\ 2$ (a) and $U/M=4,\ 6,\ 8,\ 10$ (b). Other parameters are the same as in Fig.~\ref{fig-Nk}.}
\label{fig-N}
\end{figure}

In order to compare to local probing experiments and more clearly resolve the impurity resonance structure, we are also interested in the integrated DOS around the Fermi energy $E_{F}$. For this purpose, we study the properties of $\mbox{DOS}(\omega)=-\im\sum_\bfk\bfG^r(\bfk,\omega)/\pi$. Assuming $|M_+|\ll|M_z|\ll D_c$, there is only a weak angular dependence in the momentum summation and we can perform the summation analytically, giving
\begin{align}
\bfG(z)=&
	\frac{[z-\widetilde{U}(z)]\sigma_0\!+\!\widetilde\bfM(z)\! \cdot \! \bfsigma}{4\pi(\hbar v_F)^2}
		\log\frac{|\widetilde\bfM(z)|^2-(z-\widetilde{U}(z))^2}{D_c^2}
		.
\end{align}
The resulting DOS is plotted in Fig.~\ref{fig-N} for different ratios $U/M$. In the case of vanishing $U$, Fig.~\ref{fig-N}(a), the DOS retains the linear dispersion of the surface states and has a distinct energy gap centered around the Dirac point, as is expected for a finite concentration of purely magnetic impurities \cite{SCZhangimp, Abanin11}. 
Slightly increasing $0<U/M\lesssim1$, there is an overall energy shift of the spectrum towards a hole doped state, but more importantly, the size of the energy gap also decreases. The reduced gap is a direct consequence of the impurity resonance from the potential scattering moving toward lower energies and effectively crowding out the gap. Already for $U/M = 2$, a very realistic value for the potential scattering of a magnetic impurity atom (see Concluding remarks), there is a clear resonance peak visible in the  low-energy spectrum and its tail has already started to fill up the energy gap. Thus, we again see clearly how including a realistic potential scattering term lifts the magnetically induced gap and, as a consequence, the DOS instead tends towards that of a potential impurity \cite{Biswas10, Black-Schaffer12imp, Teague12, Alpichshev_PRL12, JFransson14}. 
This offers a simple explanation to recent experimental studies, which have found no significant difference between magnetic and non-magnetic surface impurities \cite{Bianchi11, Valla12}.
We also note how the impurity resonance at intermediate $U$ values tends to split up in two peaks, a feature which has also been reported experimentally \cite{Chen10}.

\section{3D lattice model}
The continuum model results demonstrate the importance of including the potential scattering contribution for magnetic impurities, since the resulting impurity resonance easily fills the magnetically induced gap. In fact, a realistic impurity can easily provide a potential scattering perturbation ($\gtrsim 1$~eV) exceeding that of the bulk energy gap in a TI ($\sim 0.3$~eV). This also raises the question if a continuum surface model, with an infinitely large bulk gap, accurately captures the low-energy properties of TI surface impurities. 
We therefore also study magnetic surface impurities with finite potential scattering in a full 3D lattice model. This lattice calculation excellently complements the continuum model, as it not only includes a finite bulk gap, but also captures the nonzero penetration of the surface states into the bulk of the TI. Moreover, we only access the behavior of Dirac delta impurities in the continuum model, whereas the lattice model naturally allows for non-singular impurities.

More specifically, we use a simple tight-binding model of a TI, which consists of $s$-orbitals arranged on a diamond lattice \cite{Kane_PRL07, Kane_PRB07}:
%
\begin{align}
\label{eq:H0}
\Hamil_{\rm latt} = \! \! \! \! \sum_{\langle i,j\rangle,\sigma} \! \! \! (t+\delta t_{ij}) c^\dagger_{i\sigma}c_{j\sigma} +
\frac{4i\lambda}{a^2} \! \! \! \! \! \sum_{\langle \langle i,j\rangle \rangle,\sigma\sigma'} \! \! \! \! \! c^\dagger_{i\sigma} {\bm \sigma} \! \cdot \! ({\bf d}^1_{ij}\times {\bf d}^2_{ij}) c_{j\sigma'}\!.
\end{align}
Here $c_{i\sigma}$ is the annihilation operator on site $i$ in the lattice with spin-index $\sigma$,  $t$ is the nearest neighbor hopping, $\lambda$ is the spin-orbit coupling, $\sqrt{2}a$ is the cubic cell size, and ${\bf d}_{ij}^{1,2}$ are the two bond vectors connecting next-nearest neighbor sites $i$ and $j$. This system becomes a strong TI with a single Dirac surface cone when setting $\delta t_{ij} = 0.25t$ for one of the nearest neighbor directions not parallel to (111) \cite{Kane_PRL07}. To access a surface, we construct a slab in the (111) direction with ABBCC ... AABBC stacking termination. To avoid a hybridization gap between the two slab surfaces, we use eight lateral unit cells, each consisting of six atomic layers. Finally, we set $t = 2$ and $\lambda = 0.3t$, which gives $\hbar v_F\approxeq 1$ for the surface states, the same as in $\Hamil_{\rm surf}$.

\begin{figure}[t]
\begin{center}
\includegraphics[width=0.99\columnwidth]{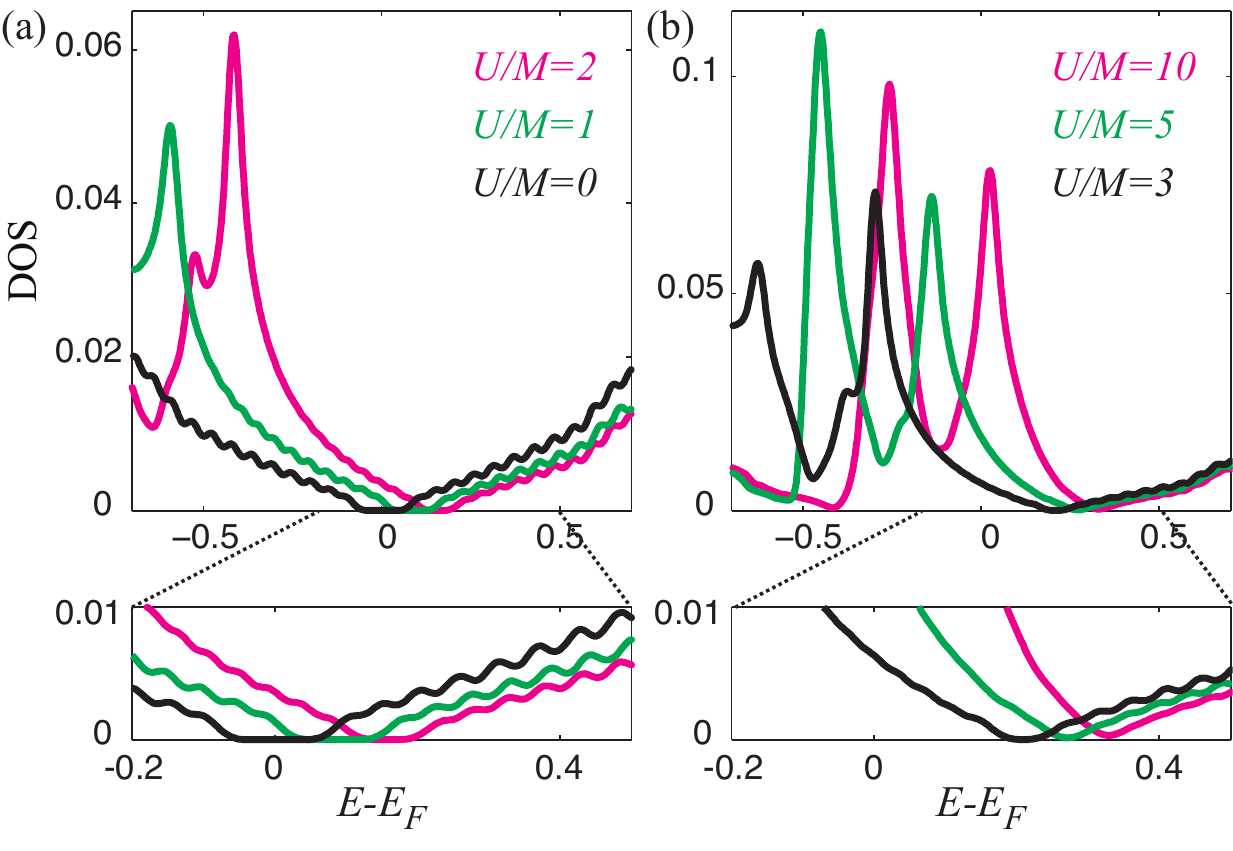}
\end{center}
\caption{(Color online) Evolution of the surface local DOS (per energy and area unit) maximally away from the impurity as a function of the impurity potential $U/M = 0, 1, 2$ (a) and $U/M = 3, 5, 10$ (b). Here $n = 4$ and $M = 4$. 
}
\label{fig:Urun}
\end{figure}
%
To study surface impurities, we create a rectangularly shaped surface supercell with $n$ sites in each direction, resulting in a surface area $A = \sqrt{3}n^2a^2/2$, where $a=1$ is the unit of length. We then add $\Hamil_{\rm imp} = \sum_{\sigma,\sigma'}  {\bf V} c_{B\sigma}^\dagger c_{B\sigma'}$  to the Hamiltonian, where ${\bf V}$ is the total impurity scattering potential and $B$ is one specific surface lattice site within the supercell.
We only consider single-site impurities, which model substitutional or on-top absorbed adatoms, as those most straightforwardly connects to the continuum model results. However, we do not expect any qualitative changes for extended defects. 
For the impurity, we include both potential and magnetic scattering: ${\bf V} = U\sigma_0 - J\bfS\cdot\bfsigma$. The surface states in $\Hamil_{\rm latt}$ do not have as simple a spin structure as the continuum model $\Hamil_{\rm surf}$, but we nonetheless find that a Zeeman magnetic field along the $\hat{\bf y}$-direction develops a clear energy gap at the Dirac point. We therefore use a total scattering potential ${\bf V} = U\sigma_0 + M\hat{\bf y}\cdot \bfsigma$ to study the influence of potential scattering on the magnetic impurity induced gap. 
We solve $\Hamil= \Hamil_{\rm latt} + \Hamil_{\rm imp}$ in the supercell using exact diagonalization and use the eigenstates to calculate the local DOS at each lattice site. We use a $70 \times 70$ $k$-point grid to achieve a high energy resolution capable of accurately resolving small energy gaps and a 0.01 Gaussian broadening to compensate for the finite system size.

In Fig.~\ref{fig:Urun}, we plot the local surface DOS maximally away from a $M = 4$ impurity in a $n = 4$ supercell, corresponding to $7\%$ surface impurity concentration. 
At $U = 0$, we see a clear energy gap at the Dirac point centered around the Fermi level. Slowly increasing $U$ moves the energy gap to slightly higher energies and, at the same time, very clear impurity resonance peaks start to appear at progressively lower energies. The latter causes the energy gap to shrink and it is completely filled already at $U/M \gtrsim 5$. We especially note that the gap disappears already for small values of $U$. This is very far from the unitary scattering limit, where the impurity resonance peak is firmly centered around the Dirac point \cite{Black-Schaffer12imp} and then generates such a large amount of DOS around the Fermi level, that the system becomes prone to spontaneous magnetization through a Stoner-like impurity mechanism \cite{Black-Schaffer14imp}. Instead, the energy gap in Fig.~\ref{fig:Urun} disappears due to the tail of the impurity resonance peak adding spectral weight at the Dirac point.
We also note that the impurity resonance has a clear double-peak structure, which we here attribute to bonding and anti-bonding impurity bands.
While we in Fig.~\ref{fig:Urun} plot the local DOS far away from the impurity, the energy gap is a global property and does not change throughout the surface. The impurity resonances are naturally taller closer to the impurity, but we find that the peaks are non-dispersive, and thus their influence on the energy gap is the same both near and far away from the impurity. 
Quite remarkably, all the results and trends with increasing strength of the potential scattering $U$ are very similar for the 3D lattice model in Fig.~\ref{fig:Urun} and the continuum surface model in Fig.~\ref{fig-Nk}. These include the energy gap shrinking and moving to higher energies, before finally disappearing for $U/M$ of the order of one, as well as the two-peaked impurity resonance structure. This is especially noteworthy since the lattice calculation has a bulk gap of only about $\pm 0.6$ and the impurity resonance state has been found to penetrate as deep as ten layers into the TI \cite{Black-Schaffer12imp, Black-Schaffer12imp2}. 

\begin{figure}[t]
\begin{center}
\includegraphics[width=0.99\columnwidth]{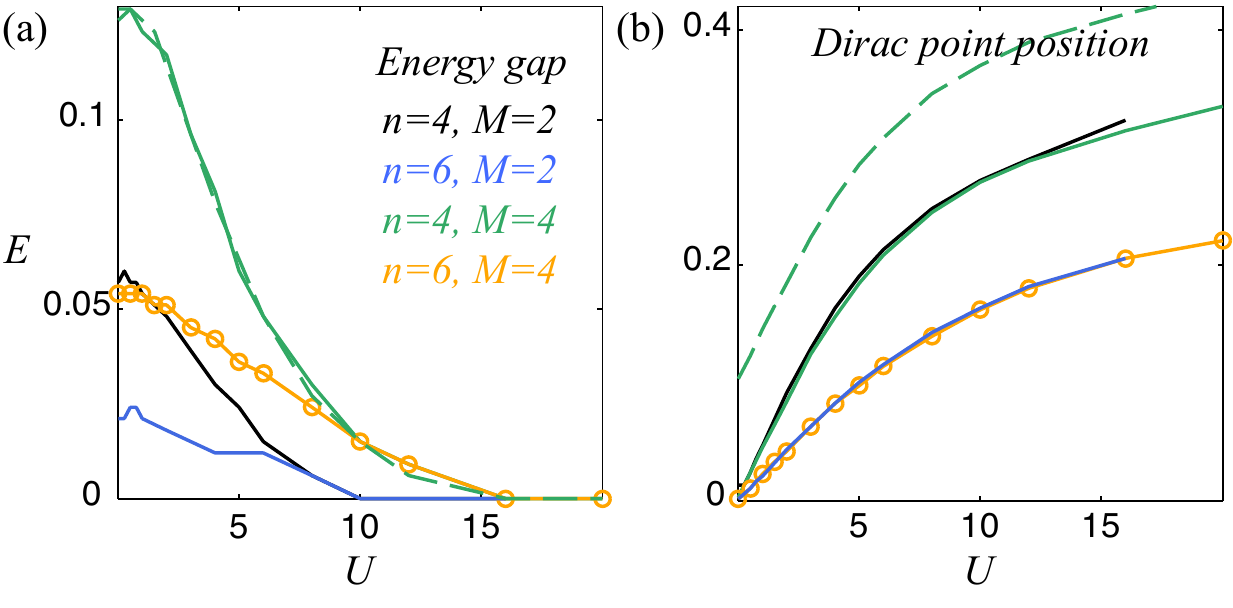}
\end{center}
\caption{(Color online) (a) Extracted energy gap and (b) position of the Dirac point (middle of the energy gap) (b) as a function of the impurity potential $U$ for supercell sizes $n$ and impurity magnetic moments $M$. Dashed line is for chemical potential $\mu = 0.1$.}
\label{fig:extgap}
\end{figure}

In order to explicitly track the influence of the potential scattering on the energy gap, we plot in Fig.~\ref{fig:extgap}(a) the extracted energy gap as function of $U$ for several different impurity concentrations and magnetic moments. 
Clearly, the energy gap is larger for larger $M$ and higher impurity concentrations, as both generate an overall larger effective Zeeman magnetic field. However, we find that the gap in all cases diminishes and is finally completely filled for very realistic strengths of the potential scattering associated with a magnetic impurity. This result is unaffected if we add a finite chemical potential $\mu$ to the Hamiltonian $\Hamil_{\rm latt}$ (dashed line). Thus both  intrinsic doping and local charge puddles \cite{Beidenkopf11} will not affect the results, if anything, non-local effects from charge puddles have been proposed to further diminish the gap \cite{Skinner13, Skinner13b}.
In Fig.~\ref{fig:extgap}(b), we finally plot the position of the Dirac point. There is a sub-linear increase in position of the Dirac point with increasing $U$, but, notably, the size of the magnetic moment does not influence the position of the Dirac point. Adding a finite $\mu$ simply shifts the Dirac point an equivalent distance.

\section{Concluding remarks}
Finite concentration of magnetic impurities, where a Zeeman magnetic field is produced as a collective effect \cite{SCZhangimp, Abanin11}, has been considered to be a promising pathway for gap opening and thus functionalizing the TI surface. 
We have here shown that by also including the potential scattering, present for all impurities, the magnetically induced gap can be completely filled. Potential scattering in Dirac materials is known to give rise to impurity resonances \cite{DMreview}, and we find that the tail of their spectral weight easily fills up the energy gap.  Most strikingly, we find that the energy gap is completely filled for a ratio between the potential and magnetic scattering contributions as low as 1 -- 10, which is well within the range of expected values for the potential (1 -- 10 eV) and magnetic (0.1 -- 1 eV) scattering  for magnetic dopants in TIs. 
The remarkably close alignment between results from a continuum surface model and a 3D tight-binding lattice model makes it possible to rule out modeling deficiencies. Our results therefore provide new crucial understanding and could offer to resolve the seemingly contradictory experimental situation as to if magnetic impurities induce an energy gap in the TI surface states.

\acknowledgments{We acknowledge useful discussions with J.~C.~Davis, N.~Nagaosa and Q.-K.~Xue. This work was supported by the Swedish Research Council, the G\"oran Gustafsson Foundation, the Knut and Alice Wallenberg Foundation, and the European Research Council under the European Union's Seventh Framework Program (FP/2207-2013)/ERC Grant Agreement No. DM-321031. Work at Los Alamos was supported by the US DOE Basic Sciences for the National Nuclear Security Administration of the US Department of Energy under Contract No. DE-AC52-06NA25396.}


\end{document}